\documentclass[a4paper,12pt]{article}
\usepackage[dvipsnames]{xcolor}
\usepackage[utf8]{inputenc} 
\usepackage[hidelinks]{hyperref}
\usepackage{float}
\usepackage{amssymb,amsmath,amsthm}
\usepackage{amsfonts}

\usepackage[draft,colorinlistoftodos, textwidth=3cm, textsize=scriptsize]{todonotes}

\pdfoutput=1
\usepackage{graphicx}
\usepackage{graphicx, float, tabularx, booktabs}
\usepackage{color}
\usepackage{enumerate}
\usepackage[american]{babel}
\usepackage{stackrel}

\usepackage[numbers, compress]{natbib}
\usepackage{natbib}
\usepackage{slashed}
\usepackage{amsfonts}
\usepackage{epsfig}
\usepackage{graphicx, tabularx}

\newcommand{\MP}{M_\mathrm{P}}
\newcommand{\be}{\begin{equation}}
\newcommand{\ee}{\end{equation}}
\newcommand{\ba}{\begin{eqnarray}}
\newcommand{\ea}{\end{eqnarray}}

\vspace{-0.1cm}
\renewcommand{\thefootnote}{\fnsymbol{footnote}}

\newcommand{\gsim}{\mathrel{\hbox{\rlap{\lower.55ex \hbox {$\sim$}}
                   \kern-.3em \raise.4ex \hbox{$>$}}}}
\newcommand{\lsim}{\mathrel{\hbox{\rlap{\lower.55ex \hbox {$\sim$}}
                   \kern-.3em \raise.4ex \hbox{$<$}}}}

 \usepackage{graphicx}

\usepackage{amsmath}

\newcommand{\bw}{\begin{widetext}}
\newcommand{\ew}{\end{widetext}}

\newcommand{\inm}[1]{\ensuremath{\text{#1}}}
\newcommand{\diag}{\ensuremath{ \text{diag} }}
\newcommand{\lz}{\ensuremath{ {l_\inm{0}} }}

\newcommand{\EpMod}{\ensuremath{ \mathfrak{T} }}
\newcommand{\pr}{\ensuremath{ {p_\inm{r}} }}
\newcommand{\pt}{\ensuremath{ {p_\inm{t}} }}

\newcommand{\der}{\ensuremath{ d }}

\def\ber{\begin{eqnarray}}
\def\eer{\end{eqnarray}}

\begin{document}

\begin{center}
{\Large \textbf{Charged black holes from T-duality}}
\end{center}

\vspace{-0.1cm}

\begin{center}
Patricio Gaete$^{a}$\footnote{%
E-mail: \texttt{patricio.gaete@usm.cl} }, Kimet Jusufi$^{b}$\footnote{%
E-mail: \texttt{kimet.jusufi@unite.edu.mk } }   and Piero Nicolini$^{c,d,e,f}$\footnote{%
E-mail: \texttt{nicolini@fias.uni-frankfurt.de} }

\vspace{.6truecm}

\emph{\small  $^a$Departamento de F\'{i}sica and 
Centro Cient\'{i}fico-Tecnol\'ogico de Valpara\'{i}so-CCTVal, Universidad T\'{e}cnica Federico Santa Mar\'{i}a,\\[-0.5ex]
\small  Valpara\'{i}so, Chile}\\[1ex]

\emph{\small  $^b$Physics Department, State University of Tetovo, \\[-0.5ex]
\small Ilinden Street nn, 1200,
Tetovo, North Macedonia}\\[1ex]

\emph{\small  $^c$Center for Astro, Particle and Planetary Physics, \\[-0.5ex]
\small New York University Abu Dhabi, 
\small  Abu Dhabi, UAE}\\[1ex] 

\emph{\small  $^d$Frankfurt Institute for Advanced Studies (FIAS),\\[-0.5ex]   Frankfurt am Main, Germany}\\[1ex]

\emph{\small  $^e$Institut f\"ur Theoretische Physik,\\[-0.5ex]  
Goethe-Universit\"at, Frankfurt am Main, Germany}\\[1ex]

\emph{\small  $^f$Dipartimento di Fisica,\\[-0.5ex]
\small Università degli Studi di Trieste, Trieste, Italy}\footnote{Present address}\\[1ex]

\end{center}
\begin{abstract}
\noindent{\small  
In this paper, we present a family of regular black hole solutions in the presence of charge and angular momentum. We also discuss the related thermodynamics and we comment about the black hole life cycle during the balding and spin down phases.
Interestingly the static solution resembles the Ayón-Beato--García spacetime, provided the T-duality scale is redefined in terms of the electric charge, $l_0\to Q$.
The key factor at the basis of our derivation is the employment of Padmanabhan's propagator to calculate static potentials. Such a propagator encodes string T-duality effects. This means that the regularity of the spacetimes here presented can open a new window on string theory phenomenology. 
}
\end{abstract}

\renewcommand{\thefootnote}{\arabic{footnote}} \setcounter{footnote}{0}

\section{Introduction}

After years of intense activity, research in quantum gravity is still in a sort of scientific limbo. Despite some progress in
the field, we still have poor understanding of the Universe at very short scales. This situation is probably due to three main problems we are currently suffering from: an excess of competing formulations, a paucity of phenomenological predictions and, most notably, a worrying absence of experimental data \cite{Nicolai13}.  

On the positive side, there is at least one thing we have fully understood. Classically curvature is the spacetime response for the presence of a certain amount of mass-energy. This is, however, just one half of the full story. 
The ability of discerning points over a manifold  depends on the resolution of a measuring device.  Quantum mechanically the resolution is related to the energy of the probe at our disposal. Such an energy cannot infinitely grow, as it is too often tacitly assumed. In the vicinity of the Planck energy, the very notion of Compton wavelength breaks down. Both spacetime and matter start being dominated by  wild quantum fluctuations.  Eventually the huge concentration of mass-energy collapses into a black hole\footnote{An alternative scenario for the end stage of the collapse is offered by the fuzzball proposal \cite{Mat05}.}. This occurrence de facto sets the Planck length as the lower bound to the accuracy of any measurement \cite{tHo90,DFG11}. 
For energies close to the Planck scale, it has been shown that a fractal can faithfully represent the foamy structure of spacetime \cite{AAS97,AJL05,MoN10}. Therefore,  \emph{fractalization} can be regarded as a new spacetime indicator. It is the response of any increase of resolution over the spacetime, much in the same way of what curvature does for  mass-energy.

The details of the aforementioned collapse into a Planckian black hole are in general not known.  The formation of an event horizon is, however, supported by studies of extreme energy scattering.  Planckian strings and black holes share important properties, e.g. nature of decay process \cite{tHo90} and correspondence between  states \cite{HoP97}. Furthermore, black holes are necessary to explain how the system classicalizes. In the above measuring experiment a further increase of energy will not lead to any improvement of the resolution. The additional energy  would rather increase the length scale of the system, that is now governed by its gravitational radius.  Ultimately, macroscopic length scales will be probed in the trans-Planckian  regime \cite{DFG11}. String theory captures such a feature too \cite{ACV87,ACV88,ACV89}. Trans-Planckian string scattering discloses a modified form of the uncertainty relation,
\ba
\Delta x\sim \frac{\hbar}{\Delta p}+\alpha^\prime \Delta p
\label{eq:gup}
\ea
known as generalized uncertainty principle (GUP). 

At first sight, fractalization can be interpreted as a nasty effect. The new term in \eqref{eq:gup}  increases the overall uncertainty. Even in the limit $\hbar\to 0$, $\Delta x$ is non-vanishing. Being $\alpha^\prime\sim G$, fractalizazion is a genuine quantum gravity phenomenon that shows up also in the absence of conventional quantum fluctuations. It is, however, possible to take advantage of an augmented uncertainty. By differentiating \eqref{eq:gup} one finds $\Delta x\geq \Delta x_\mathrm{min}\sim ({\alpha^\prime})^{1/2}$, namely a natural ultraviolet cutoff for any field theory, including general relativity.

Along this line of reasoning, quantum gravity research has undergone a ``fork''. In parallel to research on the foundations, there is now a vivid interest in  exploring the repercussion of fractalization, the only  quantum gravity ingredient we know how to implement. A notable example is offered by regular black holes in noncommutative geometry \cite{NSS06,ANS++07,Nic09}. The latter is another byproduct of string theory similar to the GUP. 

The removal of the singularity with a string induced effect like noncommutative geometry provided one of the first theoretical umbrellas for the already existing literature on regular black holes (e.g. \cite{Dym92,Bro01,MbK05,Hay06}). Due to their non-local nature, noncommutative geometry black holes have also propelled further investigations in  non-local gravity \cite{MMN11}. The latter includes deformations of the Einstein-Hilbert action that can reproduce effects of the GUP \cite{IMN13,KKM+19,CMN15,CMMN20}, unparticle physics \cite{GHS10} and the recently proposed gravity self-completeness paradigm \cite{DvG10,NiS14}. It is interesting to notice, that non-local gravity  softens the curvature singularity. The latter can, however, be improved only  upon certain conditions \cite{Nic12}.

In such a plethora of proposals, there is a mechanism that stands out. Even if \eqref{eq:gup} is only the approximation of a more complex relation containing higher order terms in $\Delta p$,  it already reveals a sort of symmetry between the ``world'' below $\Delta x_\mathrm{min}$ and above it. Such a feature emerges in a neat way  in string theory and is called T-duality. It is, however, a general property that have been noticed by  Padmanabhan \cite{Pad97,Pad98}. The spacetime reaction to the presence of a field manifests itself in a duality of the path integral.  Contributions to the integral turn to be  invariant under the exchange of the infinitesimal path length $ds\to 1/ds$. This determines the introduction of a  ``zero-point length'', $\lz$, in the field propagator, whose momentum space representation reads:
\begin{equation}
G(k)= -\frac{\lz}{\sqrt{k^2+m^2}}\, K_1 \left(\lz \sqrt{k^2+m^2}\right),
\label{eq:paddyprop}
\end{equation}
where $K_1(x)$ is a modified Bessel function of the second kind.
The above expression has three important features: 
 \begin{enumerate}[i)]
\setlength{\itemsep}{-\parsep}%
\item It is exponentially convergent at high $k$, and it correctly reproduces the standard propagator at low $k$; 
\item  It captures the good properties of both the GUP (transparent particle-black hole duality) and noncommutative geometry (non polynomial ultraviolet convergence);
\item It has a model independent  form, since it  has been derived in an alternative way also in the context of string theory   \cite{SSP03,SpF05,FSP06}.
\end{enumerate}
Recently there have been two important applications of \eqref{eq:paddyprop}. 
The first one has been the derivation of a regular spacetime geometry generated by a static, spherically symmetric, neutral source \cite{NSW19}.
The surprising feature is that the  zero point length coincides with the charge of the Bardeen solution $g\to\l_0$. Therefore, the spacetime, being formally  the same, is everywhere regular:
\ba
-g_{00}=1-\frac{2Mr^2}{\left(r^2+l_0^2\right)^{3/2}}.
\label{eq:T-neutral_metric}
\ea
The second application, has been the formulation of a finite electrodynamics governed by a T-dual photon field propagator \cite{GaN22,Mon22}. The corresponding electrostatic potential has been found to be $V\sim 1/r^{\mathbb{D}-3}$. The effective dimension $\mathbb{D}$ is a continuous number depending on the radial distance $r$. This confirms that T-duality introduces fractalization. Being $3\leq \mathbb{D}\leq 4$, such a fractalization guarantees the sought ultraviolet regularity.

Given such a background, it is natural to complete the program of black hole metrics modified by T-duality. The new effects are expected to be visible at a scale $\sim 1/l_0$, that can be anywhere between the 10 TeV and the Planck energy.
For solar masses, the parameter ordering is $l_0/GM_\odot < 10^{-23}$. With the current experimental accuracy, it is therefore unlikely to find measurable effects for stellar black holes or heavier ones. Effects are certainly important for smaller black holes, whose formation is explained by means of a de Sitter space quantum decay in the early Universe \cite{MaR95,BoH96,MaN11}. 

The neutral solution \eqref{eq:T-neutral_metric} is the ground state in parameter space. It does not suffer from quantum back reaction and it predicts  a stable cold remnant as end stage of the evaporation. Conversely, charged and rotating solutions are transient because they decay via both Hawking and Schwinger effects \cite{Gib75,Pag06}.  For microscopic black holes the discharge is almost instantaneous \cite{Nic18}. 
The study of charge and spin effects is, however, relevant for the ``balding'' and ``spin down'' phases of the black hole life cycle. In order words, the neutral solution can be either the direct product of the quantum decay of de Sitter space or/and the final configuration of a decaying black hole with charge and spin. In addition, a short decay time  is not an issue, since it is comparable with the duration the pre-inflationary era.  The neutral as well as  the charged and spinning black holes are expected to have populated the early Universe within a primordial soup of exotic objects.

Before outlining the content of the paper, it is necessary to make a remark. In the following sections, charged black hole solutions will be derived by assuming a T-duality deformation of Maxwell electrodynamics. This necessarily represents a toy model. We recently noticed that linear T-duality  electrodynamics is not realized in nature \cite{GaN22}. One cannot separate the non-linear regime from that of the onset of T-duality effects. Given the extent and complexity of the full problem, we aim to address the full scenario in a forthcoming publication. For now, we aim to focus on the understanding of the preliminary features that can show up with metrics beyond the neutral case.

The paper is organized as follows. In Section \ref{sec:staticcase}, we present the charged static black hole solution in the presence of T-duality effects for the Einstein-Maxwell system. In Section \ref{sec:thermodynamics}, we analyze the related thermodynamics and in Section we present the stationary metric \ref{sec:stationary}.

\section{Field equations and static spacetime}
\label{sec:staticcase}

The effects of T-duality have been investigated in the literature, by considering modified  field actions. For instance, the propagator \eqref{eq:paddyprop} has been introduced  in flat space electrodynamics, by assuming an action term of the kind
\begin{equation}
- \frac{1}{4}F_{\mu \nu }\, {\cal O} F^{\mu \nu }, \label{eq:fined}
\end{equation}
where 
\begin{equation}
\mathcal{O} = {\left[ {{l_0}\sqrt \Delta\,\,  {K_1}\left( {{l_0}\sqrt \Delta  } \right)} \right]^{ - 1}}, \label{NLMaxwell10}
\end{equation}
with $\Delta  \equiv {\partial _\mu }{\partial ^\mu }$.  The non-local operator $\cal{O}$ equals the unity for small arguments, while introduces an exponential damping term for $l_0\sqrt{\Delta}\gg1$.  The resulting field equation can be either written in terms of a non-local field strength tensor $\mathfrak{F}_{\mu \nu }=\mathcal{O}^{1/2}F_{\mu \nu }$ or, equivalently, in terms of the standard $F_{\mu \nu }$ coupled to a non-local current density $\mathfrak{J}^\mu=\mathcal{O}^{-1} J^\mu$ \cite{GHS10,GaN22,FNP17}. 
Along the same line of reasoning,  T-dual effects for the gravitational field  can  be derived by solving Einstein equations coupled to a non-standard energy momentum tensor \cite{NSW19,SSN06}.

For gravity and electrodynamics coupled in a system of equations, the situation is more complicated.  One starts from an action \cite{Nic18}
\begin{equation}
S=\frac{1}{2\kappa}\int \mathfrak{f}\left(R, \Box, \dots\right)\sqrt{-g}\ d^4 x + \int \mathfrak{L}\left( m_0 , F^2, \Box, \dots \right)\sqrt{-g}\ d^4 x
\label{eq:fullaction}
\end{equation}
with $\kappa=8\pi$, $\Box=\nabla_\mu \nabla^\mu$ and $\dots$ stand for higher derivative terms. Eq. \eqref{eq:fullaction} has to match the classical Einstein-Maxwell theory in the low energy/weak curvature limit $R\sim\Box\sim F^2\ll \MP^2$. The matter Lagrangian $\mathfrak{L}$ contains both the gauge field term $F^2$ coupled to gravity and a ``bare mass'' term $m_0$.  As a result, one writes:
\begin{equation}
\mathfrak{L}\left(m_0, F^2, \Box, \dots \right)\sqrt{-g}=-m_0\int d\tau \delta\left(x-x(\tau)\right)-\frac{1}{4}F_{\mu \nu }\, {\cal O}(\Box) F^{\mu \nu }\sqrt{-g}.
\label{eq:mattlagr}
\end{equation}
Variations with respect to the gauge field and the metric generate the T-duality deformed gravity-electrodynamics system of equations. Such a system has two fulfill two limits in which gravity and electrodynamics are non longer coupled, but they still maintain their non-local character. Such limits are: 
\begin{enumerate}[i)]
\setlength{\itemsep}{-\parsep}%
\item For $F^2 =0$, one must find the non-local Einstein equations 
\begin{equation}
R_{\mu\nu}-\frac{1}{2}g_{\mu\nu} R= \kappa \ \mathfrak{T}_{\mu\nu}^\mathrm{bare}
\label{eq:nlee}
\end{equation}
where $\mathfrak{T}_{\mu\nu}^\mathrm{bare}={\cal O}^{-1}(\Box) T_{\mu\nu}(m_0)$ is an effective energy momentum tensor,  $T_{\mu\nu}(m_0)$ results from the variation of  the term depending on $m_0$ in \eqref{eq:mattlagr}, while Einstein tensor is unmodified;
\item For $R=0$, one ends up with a finite electrodynamics in flat space described by the action \eqref{eq:fined}.
\end{enumerate}
Before proceedings, one has to make two further remarks. 
First, the profile of the Lagrangian $\mathfrak{f}\left( R, \Box, \dots \right)$ generating \eqref{eq:nlee} has been studied by several authors (see e.g \cite{MMN11,GHS10,Kra87,Tom97,Bar03,Dva07,Mof11,Mod12,BiM12}) and it is not unique. 
Second, following the method used to derive charged noncommutative black holes \cite{ANS++07}, we will assume that, up to some terms coming from $\mathfrak{f}\left( R, \Box, \dots \right)$, the energy momentum tensor can be written as
\begin{equation}
\mathfrak{T}_{\mu\nu}=\mathfrak{T}_{\mu\nu}^\mathrm{bare}+\mathfrak{T}_{\mu\nu}^\mathrm{em}
\label{eq:generalemt}
\end{equation}
where 
\begin{equation}
 \EpMod_{\mu \nu}^\mathrm{em}=\frac{1}{4 \pi}\left(\mathfrak{F}_{\mu \sigma}{\mathfrak{F}_{\nu}}^{\sigma}    -\frac{1}{4}g_{\mu \nu}\mathfrak{F}_{\rho \sigma}\mathfrak{F}^{\rho \sigma}\right)
\end{equation}
and $\mathfrak{F}_{\mu \nu}= {\cal O}^{1/2}(\Box) F^{\mu\nu}$.
The above energy momentum tensor will enter the non-local Einstein equations, similarly to what done in \eqref{eq:nlee} for the neutral case.  To justify the above procedure, we proceed as follows.

The general solution of gravity field equations in the presence of a static, spherically symmetric source  can be written in a compact form as
\begin{equation}
\label{eq:lineElem}
\der s^2
= g_{00}\, dt^2 +g_{rr}\, dr^2 +r^2(d\theta^2+\sin^2\theta d\phi^2)
\end{equation}
with
\begin{equation}
\begin{split}
g_{rr}^{-1}(r)
&= \left(1 - \frac{2 m(r)}{r}\right)
\end{split}
\end{equation}
and 
\begin{equation}
m(r)=-4\pi\int_0^r dr^\prime (r^\prime)^2\ \EpMod_0^0(r^\prime).
\label{eq:cumulativemass}
\end{equation}
One of the most formidable aspects of the above relation is that the integration measure is the same as in flat space. In practice, gravity couples to the Newtonian mass, a fact that drastically simplifies the derivation of the full solution \cite[p. 126]{Wal84}. Modifications with respect to general relativity are encoded in $m(r)$, whose profile is determined by T-duality. 
 The flat space calculation also allows for the  derivation of the T-dual modified Maxwell field strength  $F^{\mu\nu}$ coupled to a non-standard current $\mathfrak{J}^\nu={\cal O}^{-1}(\Box)J^\nu$. In such a way, one can represent $\EpMod_{\mu \nu}^\mathrm{em}$ in terms of $F^{\mu\nu}$, rather than in terms of the non-local tensor $\mathfrak{F}_{\mu \nu}$. 
 
 In conclusion, the problem of deriving the static spacetime is reduced to solving the following system 
\begin{eqnarray}
&& R_{\mu\nu}-\frac{1}{2}g_{\mu\nu} R= \kappa \ \left(\mathfrak{T}_{\mu\nu}^\mathrm{bare}+\mathfrak{T}_{\mu\nu}^\mathrm{em}\right)\\
&&\nabla_\mu F^{\mu\nu} =4\pi \ \mathfrak{J}^\nu, 
\label{eq:nlmaxwell}
\end{eqnarray}
with $\mathfrak{T}_{\mu\nu}^\mathrm{bare}={\cal O}^{-1}(\Box) T_{\mu\nu}(m_0)$, $\mathfrak{J}^\mu=\mathcal{O}^{-1}(\Box) J^\mu$ and
\begin{equation}
 \EpMod_{\mu \nu}^\mathrm{em}=\frac{1}{4 \pi}\left(F_{\mu \sigma}{F_{\nu}}^{\sigma}    -\frac{1}{4}g_{\mu \nu}F_{\rho \sigma}F^{\rho \sigma}\right).
 \label{eq:emtem}
\end{equation}
In the process, one can neglect any coupling between the gauge field and non-local terms coming from the gravity Lagrangian  $\mathfrak{f}\left( R, \Box, \dots \right)$, since they do not affect $m(r)$.  It is, however, necessary to comment about the relevance of such non-local terms, that generate a correction to the above energy momentum tensor. We will expand the discussion about this point in the next section.

\subsection{T-duality improved line element}

The cumulative mass $m(r)$ 
requires the integration of the density of the bare mass and the energy stored in the electromagnetic field:
\begin{equation}
\EpMod_0^0=-\rho(r)=-\rho_\mathrm{bare}(r)-\rho_\mathrm{em}(r).
\label{eq:tnaughtnaught}
\end{equation} 
The conservation of the energy momentum implies the presence of non-vanishing pressure terms, namely ${\EpMod^\mu}_\nu = \diag\left(-\rho(r),\ \pr(r),\ \pt(r),\ \pt(r)\right)$. 

The above density and pressure functions have to describe the conventional  electrovacuum  at large distances to guarantee the matching with the  Reissner-Nordström geometry. Conversely, at short distances,  T-duality sets a scale $l_0$ at which quantum effects become relevant. In such a region,  energy conditions are violated, 
opening the  possibility of evading the singularity theorems \cite{Pen65} to obtain a regular solution.  
The physical picture is that the quantum fluctuations associated to $l_0$ counteract the gravitational pull and prevent the gravitational collapse. The average of such  fluctuations over a volume $\sim l_0^3 $ give rise to an anti-gravity region, filled with a fluid whose density and pressure are finite.

The solution of the system of equation requires an equation of state for pressure terms. The relation $\pr(r)=-\rho(r)$ is the most natural choice, because it implies the Schwarzschild like condition for the metric coefficients, $g_{00}=-g_{rr}^{-1}(r)$. As a result the shape function $m(r)$ becomes the only unknown. At short distances, the regularity of the source implies 
\begin{equation}
\pr(r)=-\rho(r) \stackrel[r\lesssim l_0]{}{\leadsto} \pr(0)=-\rho(0)\approx \mathrm{const.}\leadsto m(r)\propto  r^3.
\end{equation}
In practice, the aforementioned anti-gravity region is a ball of de Sitter spacetime at the origin, whose cosmological constant  corresponds to the (regularized) quantum vacuum energy emerging from virtual particle exchange.

With the above equation of state, the energy momentum tensor can be written as
\begin{equation}
\EpMod_{\mu\nu}=-\rho(r)\ g_{\mu\nu}+t_{\mu\nu},
\end{equation}
with 
\begin{equation}
t_{\mu\nu}=\diag\left(0,\ 0,\ -(r/2) \ d\rho(r)/dr,\ -(r/2) \ d\rho(r)/dr \right).
\end{equation}
The tensor $t_{\mu\nu}$ vanishes at the origin and at large distances with respect to $l_0$.  It is non-vanishing only in the middle range and breaks the isotropy  of $\EpMod_{\mu\nu}$. In addition, one can split it in two terms, namely  $t_{\mu\nu}=t_{\mu\nu}^\mathrm{bare}+t_{\mu\nu}^\mathrm{em}$. The term $t_{\mu\nu}^\mathrm{bare}$ is already included in $T_{\mu\nu}^\mathrm{bare}$, namely
\begin{equation}
T_{\mu\nu}^\mathrm{bare}=-\rho(r)\ g_{\mu\nu}+t_{\mu\nu}^\mathrm{bare}.
\end{equation}
The new term is actually $t_{\mu\nu}^\mathrm{em}$. As a result, the tensor  $\EpMod_{\mu\nu}^\mathrm{em}$ is no longer traceless. Thus, one can conclude that \eqref{eq:emtem} provides only the leading term of the energy momentum of the non-local electromagnetic field. As previously said, we have neglected the coupling of non-local terms from the gravity action with the gauge field. Such coupling depends on the scale $l_0$ and it is reasonable to expect it to break the tracelessness property of $ \EpMod_{\mu \nu}^\mathrm{em}$.  The action, whose variation generates $T_{\mu\nu}^\mathrm{bare}$,  has been derived in \cite{MaN11}. The action for $t_{\mu\nu}^\mathrm{em}$ requires additional work, due to the arbitrariness of the Lagrangian $\mathfrak{f}\left( R, \Box, \dots \right)$.  We postpone such an analysis  to another publication. As stated above, $t_{\mu\nu}^\mathrm{em}$ cannot affect the derivation of the spacetime geometry, since $m(r)$ does not depend on it.

To obtain the full solution of field equations, one has to determine  the energy densities \eqref{eq:tnaughtnaught}, by using  the T-dual propagator \eqref{eq:paddyprop}.
To this purpose, we recall that  the Newtonian potential can be calculated from 
 the inverse Fourier transform of the Feynman propagator. As a result,
\begin{equation}
V(r)=-m_0\int\frac{d^3 k}{(2\pi)^3} \ \left. G_\mathrm{F}(k)\right|_{k^0=0} e^{i\mathbf{k}\cdot\mathbf{x}}=-\frac{M}{\sqrt{r^2 + \lz^2}},
\end{equation}
where one has used, in place of the conventional $G_\mathrm{F}(k)$, the Padmanabhan's propagator \eqref{eq:paddyprop} to obtain the sought T-duality effect. 
 Then, from the Newton-Poisson equation one finds
\begin{equation}
\frac{1}{4\pi}\Delta V(r)\equiv\rho_\mathrm{bare}(r)= \frac{3 \lz^2 m_0}{4 \pi {\left(r^2 +\lz^2\right)}^{5/2}}.
\end{equation}
The above result is the generalization of the conventional Newton-Poisson equation, whose source is expressed in terms of a Dirac delta distribution. Balasin and Nachbagauer have shown that the method to obtain $\rho_\mathrm{bare}(r)$ works for the Schwarzschild metric too \cite{BaN93}. 
Indeed, classical spacetimes are weak solutions that correspond to an energy momentum tensor, written in terms of distributions rather than functions \cite{BaN94}.  

As a next step, one has to determine the energy stored in the electromagnetic field. From \eqref{eq:nlmaxwell}, one obtains
 $A_t=-Q/(\sqrt{r^2+l_0^2})$. The only  non-vanishing components of the field strength tensor are:
\begin{equation}
F_{0r}=-F_{r0}=-\frac{Q\, r}{\left(r^2+l_0^2\right)^{3/2}}.
\label{eq:staticem}
\end{equation}
As a result one has
\begin{equation}
 \EpMod_{0}^{0\ \mathrm{em}}=\frac{Q^2 r^2}{8 \pi (r^2+l_0^2)^3},
 \label{eq:estaticen}
\end{equation}
that is vanishing at the origin, decaying as $\sim 1/r^4$ at infinity, and everywhere finite.
One can then easily obtain the following exact solution
\begin{eqnarray}\notag
   -g_{00}&=&1-\frac{2m_0 r^2}{\left(r^2+l_0^2\right)^{3/2}}+\frac{5 Q^2 r^2}{8 (r^2+l_0^2)^2}\\
&+&\frac{3 Q^2 l_0^2}{8 (r^2+l_0^2)^2}-\frac{3 Q^2}{8 l_0\,r }\arctan\left(\frac{r}{l_0}\right).
\end{eqnarray}
The above spacetime is everywhere regular. 

To grasp the full meaning of the above line element, it is important to study its limits. At short scales, namely $r\ll l_0$ one finds that 
\begin{equation}
-g_{00}\approx 1- \frac{2m_0}{l_0^3} \ r^2 + O\left(r^4\right).
\end{equation}
At the origin,  the de Sitter ball depends on $m_0$ only. This is not a surprise. T-dual electrodynamics has finite static potentials \cite{GaN22}.  From \eqref{eq:staticem} one finds that  fields are linearly vanishing. The energy momentum tensor goes to zero quadratically (see \eqref{eq:estaticen}). Therefore the electrostatic contribution to $m(r)$ goes like $\sim r^5$ to zero at short scale. In practice, the mass cannot be dressed in the vicinity of the origin. In the limit $m_0\ll Q$, the spacetime is still regular but the de Sitter ball is replaced by a Minkowski region.  

At large distances $r\gg l_0$ one finds that the solution is
\begin{equation}
-g_{00}\approx 1 - \frac{2m_0+\frac{3\pi Q^2}{16 l_0}}{r}+\frac{Q^2}{r^2}+ O\left(r^{-3}\right)
\end{equation}
The above metric has to describe the standard Reissner-Nordström. This necessitates the identification of the ADM mass $M$ with
\begin{equation}
M=m_0+\frac{3\pi Q^2}{32\ l_0}.
\label{eq:admmass}
\end{equation} 
Amazingly, one discovers that the mass contains a new term proportional to the regularized self energy of the electrostatic field. In general relativity, such a term is simply neglected, because it is part of the curvature singularity.

We can use \eqref{eq:admmass} to write the metric coefficient in its final form:
\begin{equation}
-g_{00}=1-\frac{2Mr^2}{\left(r^2+l_0^2\right)^{3/2}}+ \frac{Q^2 \ r^2}{\left(r^2+l_0^2\right)^{2}}F(r)
\label{eq:lineelfinal}
\end{equation}
where
\begin{equation}
F(r)=\frac{5}{8}+\frac{3l_0^2}{8r^2}+\frac{3\pi}{16l_0}\left(r^2+l_0^2\right)^{\frac{1}{2}} \left\{1-\frac{\left(r^2+l_0^2\right)^{\frac{3}{2}}}{r^3}\left[\frac{2}{\pi}\arctan\left(\frac{r}{l_0}\right)\right]\right\}.
\end{equation}
The function $F(r)$ is a monotonic increasing function. It tends to $1$ for $r\gg l_0$ and to $3\pi/16\simeq 0.59$ for $r\ll l_0$. This means $3\pi/16< F(r)< 1$. It was already observed that the neutral solution is formally equivalent to the Bardeen metric provided one redefines the length $l_0\to Q$ \cite{NSW19}. It is very interesting to notice that for the metric \eqref{eq:lineelfinal}, the same substitution $l_0\to Q$ leads to another relevant line element, namely the spacetime derived by Ayón-Beato  and García \cite{AyG99c}, provided $F(r)=1$ everywhere. 

 The above solution \eqref{eq:lineelfinal} admits a horizon structure similar to the Reissner-Nordström one. There exists a value of the mass $M_\mathrm{ext}$ such that the black hole is in its extremal configuration, namely it has one degenerate horizon. For masses $M>M_\mathrm{ext}$, there exist an outer event horizon and a inner Cauchy horizon. For $M<M_\mathrm{ext}$, there spacetime has no horizon but just a weak, regular curvature. This case is suitable to describe a charged particle, that is too light to create an event horizon. The extremal black hole configuration plays an important role in the thermodynamics of the system, that is the subject of the next section.

\section{Black hole thermodynamics}
\label{sec:thermodynamics}

\begin{figure}
  \includegraphics[width=0.9\linewidth]{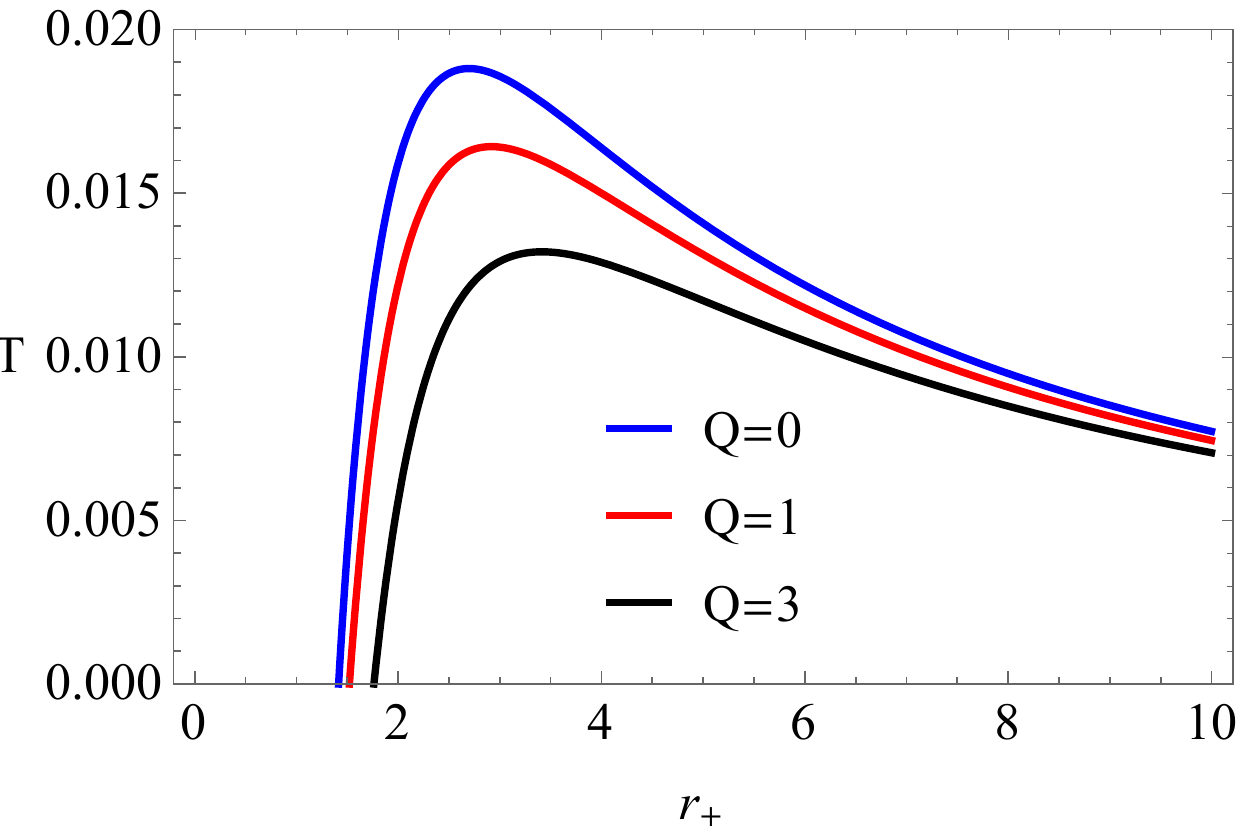}
\caption{The Hawking temperature as a function of $r_+$. We have set $M=5$ and  $l_0=1$ in Planck units. }
\label{fig:temp}
\end{figure}

To study the thermodynamics of the spacetime geometry it is convenient to express \eqref{eq:lineelfinal} in a more compact form. Along the line of what shown in \cite{Nic10}, one has to introduce a function $\mathcal{G}(r)$ such that
\begin{equation}
-g_{00}=1-\frac{2M}{r}\ \mathcal{G}(r).
\label{eq:genmetric}
\end{equation}
With the above representation, we incorporate in $\mathcal{G}(r)$ all deviations from the Schwarzschild metric, due to interactions (e.g. charge) and/or short scale quantum effects.
In our case one finds
\begin{equation}
\mathcal{G}(r)=\frac{r^3}{\left(r^2+l_0^2\right)^{\frac{3}{2}}}\left[1 - \frac{Q^2}{2M}\frac{F(r)}{\left(r^2+l_0^2\right)^{\frac{1}{2}}}\right],
\end{equation}
that in the large $r$ limit consistently reproduces the Reissner-Nordström case, $\mathcal{G}\to 1-Q^2/(2Mr)$. At this point, one can exploit the fact that the Hawking temperature reads
\begin{equation}
T=\frac{1}{4\pi r_+}\left[1-r_+\  \frac{d\mathcal{G}(r_+)/dr_+}{\mathcal{G}(r_+)}  \right]
\end{equation}
where $g_{00}(r_+)=0$ for $M\geq M_\mathrm{ext}$ with $r_+\geq r_\mathrm{ext}\geq \sqrt{2}\ l_0$. The last inequality is saturated when $Q=0$. The temperature is plotted in Fig. \ref{fig:temp}. One can see that similarly to Reissner-Nordström the temperature admits a maximum before a cooling down phase to the extremal configuration $r_\mathrm{ext}, \ M_\mathrm{ext}$. In contrast to Reissner-Nordström, such a cooling down phase persists also in the case of vanishing $Q$. This is a crucial feature that guarantees the stability of the solution also in case of sudden discharge via Hawking and Schwinger effects. The  charge simply decreases the maximum of the temperature, and increases the size of $r_\mathrm{ext}$.

One can also write the entropy of the black hole as a function of $\mathcal{G}$. It reads
\begin{equation}
S(r_+)=2\pi \int_{r_\mathrm{ext}}^{r_+} \frac{rdr}{\mathcal{G}(r)} .
\end{equation}
The integration of the above equation gives: 
\begin{equation}
S(r_+)=\frac{\mathcal{A}_+}{\mathcal{G}(r_+)}- \frac{\mathcal{A}_\mathrm{ext}}{\mathcal{G}(r_\mathrm{ext})} +\pi \int_{r_\mathrm{ext}}^{r_+} \frac{r^2 \mathcal{G}^\prime(r) dr}{\mathcal{G}^2(r)},
\label{eq:finalentropy}
\end{equation}
where $\mathcal{A}_+=\pi r_+^2$ and $\mathcal{A}_\mathrm{ext}=\pi r_\mathrm{ext}^2$ are the event horizon areas. Eq. \eqref{eq:finalentropy}  is compatible with the area law up to some short scale deformations. 

\begin{figure*}
  \includegraphics[width=0.45\textwidth]{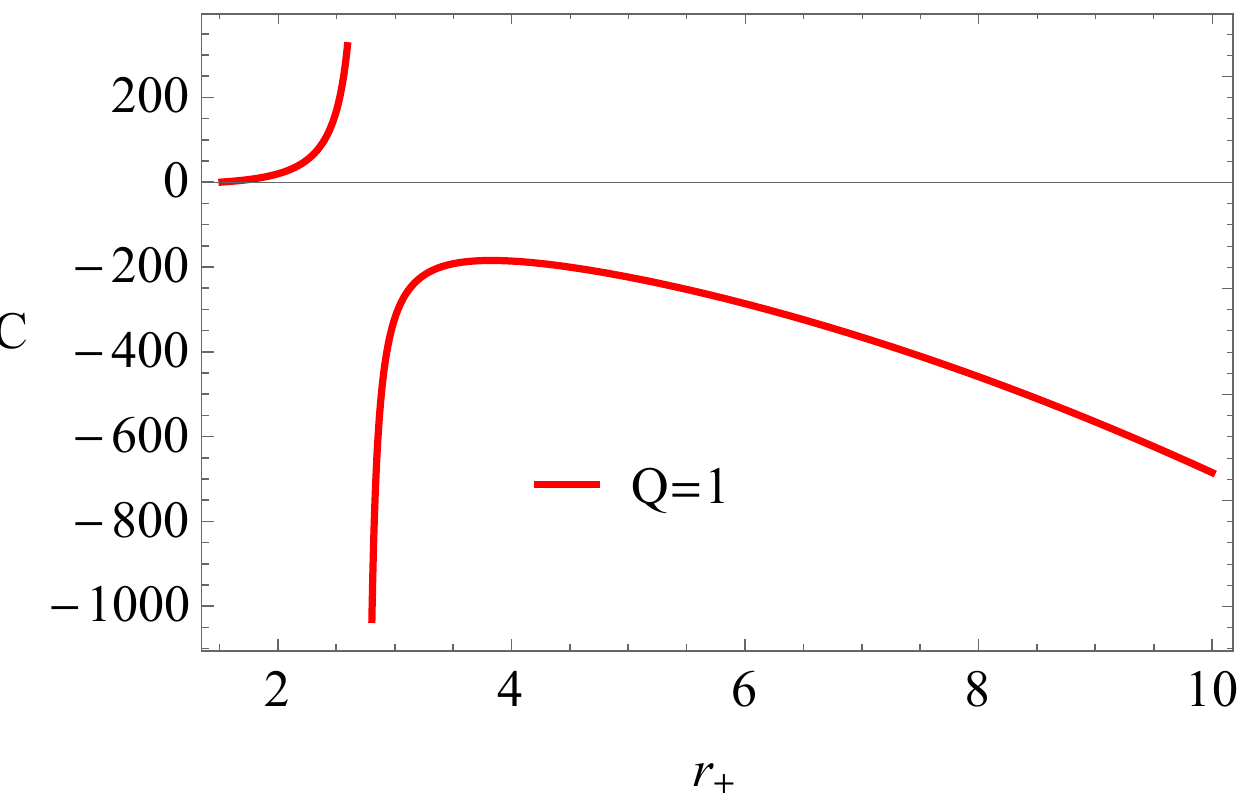}
    \includegraphics[width=0.45\textwidth]{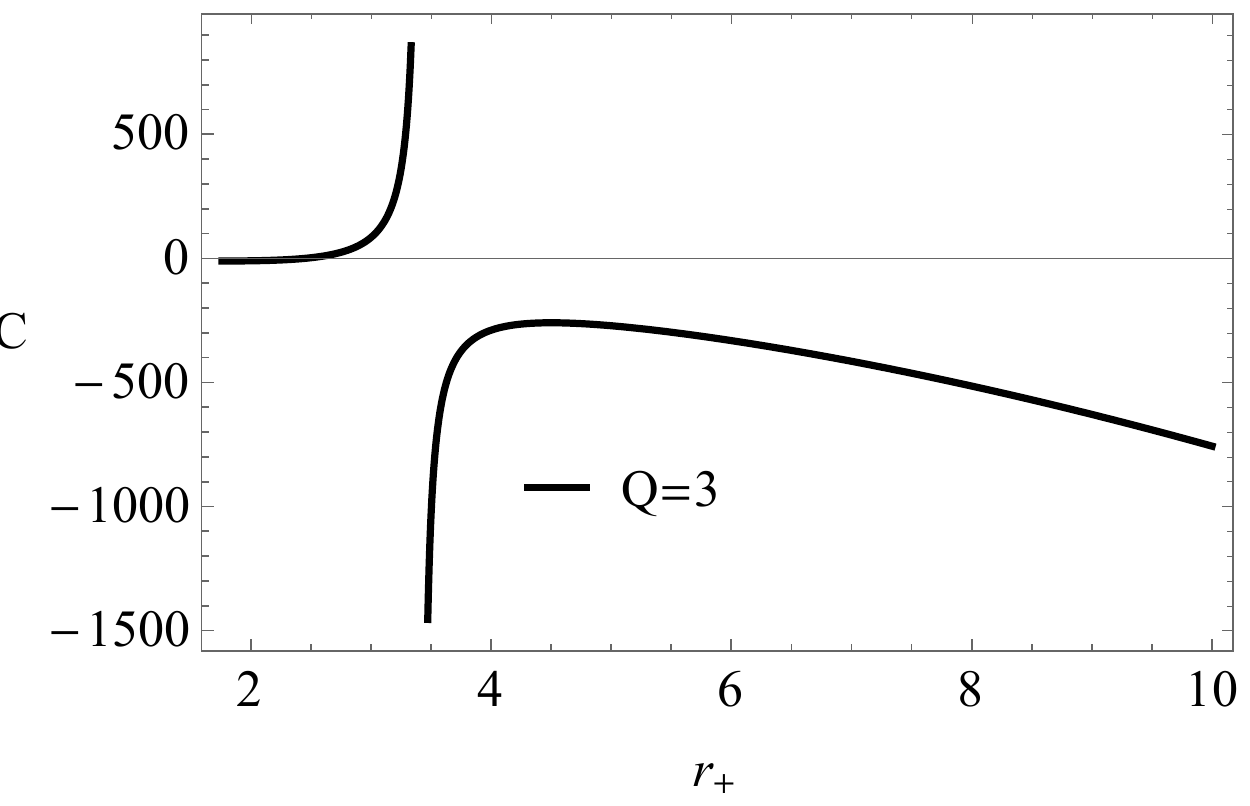}
    \caption{The plot showing the heat capacity of the black hole as a function of $r_+$. 
    We have set $M=5$ and $l_0=1$ in Planck units. }
    \label{fig:heat}
\end{figure*}

A final comment is related to the heat capacity. The general formula reads
\begin{equation}
C=\left.\frac{\partial M}{\partial T}\right|_{r_+}.
\end{equation}
We can exploit \eqref{eq:genmetric} to obtain $M=r_+/2\mathcal{G}(r)$ from the horizon equation. Then, one can write
\begin{equation}
C(r_+)=\frac{2\pi r_+}{\mathcal{G}(r_+)}\ T \left(\frac{dT}{dr_+}\right)^{-1}
\label{eq:heatcap}
\end{equation}
The full expression can be found in \cite{NiT11}. Eq. \eqref{eq:heatcap} reveals that $C$ vanishes at the extremal configuration $r_+=r_\mathrm{ext}$ and for large $r_+$. It has, however, a singularity and a sign change. In other words, the system undergoes a phase transition at the maximum temperature, i.e., for  $dT/dr_+=0$. The cooling down phase, known as black hole SCRAM\footnote{The term ``black hole SCRAM'' has been introduced by one of us to describe the cooling down  of an evaporating black hole \cite{Nic09}. 
Originally,  SCRAM is a backronym for ``Safety control rod axe man''. Enrico Fermi used it for the first time in 1942
during the Manhattan Project at Chicago Pile-1. The term is still used today for emergency shutdowns of 
nuclear reactors.}, is a stable, positive heat capacity, asymptotic approach to the extremal configuration. The phase preceding the peak of the temperature is unstable with negative heat capacity. The heat capacity is plotted in Fig. \ref{fig:heat}, that confirms the above analysis. The effect of the charge is to shift the phase transition (i.e. vertical asymptote) to larger $r_+$. 

\section{Stationary spacetime}
\label{sec:stationary}

In the phase preceding the neutral, static configuration, the black hole could have had not only charge but also angular momentum. Without entering the details of the spin down process in the presence of T-duality effect, we simply introduce the rotating black hole solution, whose vanishing momentum limit gives \eqref{eq:lineelfinal}.
%
 %
To reach the goal, we shall follow a method based on  a modification of the Newman-Janis algorithm, that does not require the complexification of the radial coordinate  \cite{Azreg-Ainou:2014pra}. To ease the notation, we introduce the function $f(r)=-g_{00}=g_{rr}^{-1}$. Then, we switch from the Boyer-Lindquist (BL) coordinates $(t,r,\theta,\phi)$ to the Eddington-Finkelstein (EF) coordinates $(u,r,\theta,\phi)$. In such a way, the metric can be decomposed using  null tetrads
	\begin{eqnarray}
		l'^{\mu}&=&\delta^{\mu}_{r},\\
		n'^{\mu}&=&\delta^{\mu}_{u}-\frac{1}{2}\mathcal{F}\delta^{\mu}_{r},\\\label{e11}
		m'^{\mu}&=&\frac{1}{\sqrt{2\,\mathcal{H}}}\left[(\delta^{\mu}_{u}-\delta^{\mu}_{r})\dot{\iota}{a}\sin\theta+\delta^{\mu}_{\theta}+\frac{\dot{\iota}}{\sin\theta}\delta^{\mu}_{\phi}\right],\\
		\overline{m}'^{\mu}&=&\frac{1}{\sqrt{2\,\mathcal{H}}}\left[(\delta^{\mu}_{u}-\delta^{\mu}_{r})\dot{\iota}{a}\sin\theta+\delta^{\mu}_{\theta}+\frac{\dot{\iota}}{\sin\theta}\delta^{\mu}_{\phi}\right],
	\end{eqnarray}
	where now the functions $f(r)$ and $h(r)=r^2$ transform according to $\mathcal{F} \to \mathcal{F}(r,a,\theta)$ and $\mathcal{H} \to \mathcal{H}(r,a,\theta)$, respectively. At this point, we utilize the  non-complexification procedure and we can drop the complexification of the radial coordinate (see for details \cite{Azreg-Ainou:2014pra}).  Finally, for the Kerr-like coordinates one can show that
	\begin{eqnarray}\notag
		ds^2 &=& -\frac{\Delta}{\Sigma}(dt-a\sin^2\theta d\phi)^2+\frac{\Sigma}{\Delta}dr^2+\Sigma\,d\theta^2 \\
		\label{metric}	&+&\frac{\sin^2\theta}{\Sigma}[a dt-(r^2+a^2)d\phi]^2, 
		\label{rotating}
	\end{eqnarray}
	with
	\begin{eqnarray}\notag
    \Delta(r)&=&r^2-\frac{2M r^4}{\left(r^2+l_0^2\right)^{3/2}}+\frac{Q^2 r^4}{(r^2+l_0^2)^2} F(r)+a^2,
\end{eqnarray}
along with
	\begin{eqnarray}
		\Sigma(r,\theta)&=&r^2+a^2\cos^2\theta,
	\end{eqnarray}
	where we note that $a=L/M$ is the specific angular momentum, $M$ is the ADM mass of the black hole.  Let us also point out that the function  $\mathcal{H}(r, \theta, a)$ is still unknown and we can find it by using the
	cross-term of the Einstein tensor $G_{r \theta}=0$, for a physically acceptable rotating solution. After solving a differential equation one can show that \cite{Azreg-Ainou:2014pra}
 \begin{equation}
	    \Sigma=\mathcal{H}=h(r)+a^2\cos^2\theta=r^2+a^2 \cos^2\theta. 
	\end{equation}
	Using the non-local Einstein equations, we can find the effective energy momentum tensor represented by $\EpMod_{\mu \nu}^\mathrm{eff}=(1/8\pi)G_{\mu\nu}$. In terms of the orthogonal basis, the  components are given as follows
\begin{eqnarray}\nonumber
\rho^\mathrm{eff} &=&\frac{1}{8\pi}\emph{e}^\mu_t\,\emph{e}^\nu_t \,\emph{G}_{\mu\nu},\quad
p_r^\mathrm{eff} =\frac{1}{8\pi}\emph{e}^\mu_r\,\emph{e}^\nu_r \,\emph{G}_{\mu\nu},\\\label{m1}
p_\theta^\mathrm{eff} &=&\frac{1}{8\pi}\emph{e}^\mu_\theta\,\emph{e}^\nu_\theta \,\emph{G}_{\mu\nu},\quad
p_\phi^\mathrm{eff} =\frac{1}{8\pi}\emph{e}^\mu_\phi\, \emph{e}^\nu_\phi\, \emph{G}_{\mu\nu}.
\end{eqnarray}

Note here that one has to find an orthogonal basis such that the Einstein field equations are satisfied. There are many possibilities, but one such orthogonal basis is the following choice
\begin{eqnarray}\label{basis}\notag
{\emph{e}}^\mu_t&=&\frac{1}{\sqrt{\mathcal{H} \Delta}}\left(r^2+a^2,0,0,a\right),\\\notag
\emph{e}^\mu_r&=&\frac{\sqrt{\Delta}}{\sqrt{\mathcal{H}
}}\left(0,1,0,0\right),\\\nonumber
\emph{e}^\mu_\theta&=&\frac{1}{\sqrt{\mathcal{H}}}\left(0,0,1,0\right),\\\notag
\emph{e}^\mu_\phi&=&\frac{1}{\sqrt{\mathcal{H}} \sin\theta}\left(a \sin^2\theta,0,0,1\right).
\end{eqnarray}

The metric \eqref{rotating} represents a regular version of the Kerr-Newman  metric, whose ring singularity is replaced by a spheroidal ``de Sitter belt''. Similar scenarios have been found also in the case of other regular black hole solutions  \cite{SmS10,MoN+10}.

\section{Final remarks}

This paper is at the confluence of two recent results in the framework of T-duality  effective field theories, namely a neutral black hole solution \cite{NSW19} and a finite electrodynamics \cite{GaN22}. The key ingredient in both cases is a string modified form of field propagators that Padmanabhan has introduced on the ground of model independent quantum gravity arguments  \cite{Pad97,Pad98}.

Accordingly, we presented a novel charged black hole solution with T-duality induced short scale  corrections. Interestingly, the new solution is regular and resembles the Ayón-Beato--García spacetime \cite{AyG99c}, provided the T-duality scale is redefined in terms of the electric charge, $l_0\to Q$. This property is complementary to what already observed for the neutral solution:  the substitution $l_0\to Q$ makes the metric equivalent to the Bardeen one.  

On the thermodynamic side, we derived the  Hawking temperature, the black hole entropy and the heat capacity. Similarly to the Reissner-Nordström case, the thermodynamics is stable in the final stage of the evaporation for the presence of a phase transition to a positive heat capacity cooling down. In contrast to the Reissner-Nordström case, the solution remains stable also in the vanishing charge limit. In practice there is no ``Schwarzschild phase'' as a resulting of the black hole ``balding phase'', during which the charge is shed. 

In the final part of the paper, we presented also the corresponding spinning solution. This is the starting point for further studies of the full scenario of the black hole life cycle that includes a ``spinning down'' phase too. 

Our results are general but we expect them to be relevant at energy scales of the order $1/l_0>10$ TeV. With the current observational accuracy it is unlikely that the proposed effects can be exposed in solar mass black holes or heavier one.
It is, however, reasonable to think such result to have an impact in the physics of the early Universe, whose evolution could have been affected by the presence of T-duality corrected black holes. We aim to investigate the robustness of such an idea   in forthcoming publications.

\section*{Acknowledgments}

One of us (P. G.) was partially supported by ANID PIA / APOYO AFB180002.
The work of P.N. has partially been supported by GNFM, Italy's National Group for Mathematical Physics.

\end{document}